\documentclass{jfm}
\usepackage{graphicx}
\usepackage{epstopdf, epsfig}

\usepackage[usenames]{color}

\newcommand{\SM}[1]{\textcolor{black}{#1}}

\shorttitle{Rheology of conductive particles in electric field}
\shortauthor{S. Mirfendereski, and J.S. Park}

\title{Rheology of dense suspensions of ideally conductive particles in an electric field}

\author{Siamak Mirfendereski\aff{1}
  \and Jae Sung Park\aff{1}
  \corresp{\email{jaesung.park@unl.edu}}}

\affiliation{\aff{1}Department of Mechanical and Materials Engineering, University of Nebraska-Lincoln,
Lincoln, NE 68588-0526, USA
}

\begin{document}

\maketitle

\begin{abstract}
The rheological behaviour of dense suspensions of ideally conductive particles in the presence of both electric field and shear flow is studied using large-scale numerical simulations. Under the action of an electric field, these particles are known to undergo dipolophoresis, which is the combination of two nonlinear electrokinetic phenomena -- induced-charge electrophoresis and dielectrophoresis. For ideally conductive particles, induced-charge electrophoresis is predominant over dielectrophoresis, resulting in transient pairing dynamics. The shear viscosity and first and second normal stress differences $N_1$ and $N_2$ of such suspensions are examined over a range of volume fractions $15\% \leqslant \phi \leqslant 50\%$ as a function of Mason number $Mn$, which measures the relative importance of viscous shear stress over electrokinetic-driven stress. For $Mn < 1$ or low shear rates, the dipolophoresis is shown to dominate the dynamics, resulting in a relatively low-viscosity state. The positive $N_1$ and negative $N_2$ are observed at $\phi < 30\%$, which is similar to Brownian suspensions, while their signs are reversed at $\phi \ge 30\%$. For $Mn \ge 1$, the shear thickening starts to arise at $\phi \ge 30\%$, and an almost five-fold increase in viscosity occurs at $\phi = 50\%$. Both $N_1$ and $N_2$ are negative for $Mn \gg 1$ at all volume fractions considered. We illuminate the transition in rheological behaviours from dipolophoresis to shear dominance around $Mn = 1$ in connection to suspension microstructure and dynamics. Lastly, our findings reveal the potential use of nonlinear electrokinetics as a means of active rheology control for such suspensions.
\end{abstract}


\section{Introduction}\label{sec:intro}
Electric-field-driven suspensions of small particles have been appreciated as a promising context for effectively manipulating the particle configuration and actively controlling the stress transfer of fluids \citep{yeh1997,velev2006,sheng2012,li2018,driscoll2019}.
Playing a crucial role in the driving mechanism of such fluids, nonlinear electrokinetic phenomena have been widely exploited in a variety of fields, among which are material science, bioengineering, and nano-and microfluidics \citep{feng2020,xuan2022}. 
In particular, a growing interest in slurry batteries containing suspended conductive particles requires a deep investigation into nonlinear electrokinetic phenomena resulting specifically from particle polarizability \citep{presser2012,soloveichik2015,tan2019,sanchez2021,heidarian2022}. However, the current understanding of utilizing the electrokinetics for tuning the suspension rheology is mainly limited to non-conductive or weakly conductive particle systems. 

For non-conductive or weekly conductive particles suspended in an insulating fluid, the classical example of nonlinear electrokinetics is dielectrophoresis (DEP), under which the particles experience dipole moments upon the application of an electric field. Resulting dipolar interactions between the particles cause them to assemble into chain/column-like structures along the field direction. Such a well-known response belongs to a popular class of smart fluids or the so-called electrorheological (ER) fluid \citep{wang2000,sheng2012}. 
As a main characteristic of the typical ER fluids, the rapid formation of the field-aligned anisotropic structures under the action of an electric field induces a dramatic viscosity enhancement, potentially leading to a phase transition from a liquid to a solid state. These reversible and controllable features make the ER fluids an attractive material of choice for a variety of applications, including valves \citep{whittle1994,choi1997}, active shock absorbers \citep{choi2007}, clutches \citep{madeja2011}, and actuators \citep{mazursky2019}. Furthermore, as relevant to technical applications of the ER fluids, their rheological response to an external shear flow has been extensively explored \citep{klingenberg1989,klingenberg1991small, bonnecaze1992yield,bonnecaze1992dynamic,qian2013}. The rheological properties are strongly related to the response of the field-induced structures to the external flow, where the elastic body-like deformation was developed at low shear rates, while rapid transient structural arrangement occurred at high shear rates \citep{parthasarathy1996,qian2013}. Hence, such rheological behaviour of the ER fluids was primarily described by the Bingham constitutive model \citep{bonnecaze1992yield}. In addition to the ER fluid, a similar rheological response can also be generated by applying a magnetic field to a suspension, which leads to a well-known concept of the so-called magnetorheological fluid \citep{von2002,vicente2011,lopez2016}.

Rheological properties of a suspension of non-conductive particles can also be modified by a completely different mechanism than the ER fluid. Upon the application of strong electric fields above the critical strength, the particles can start to undergo spontaneous sustained rotation via an effect of the so-called Quincke rotation \citep{quincke1896}. This effect arises primarily due to the symmetry breaking when the polarized cloud of ions around the particle induces a dipole anti-parallel to the field, which is the case when the charge relaxation time of the particles is greater than that of the suspending fluid \citep{lobry1999,das2013,saintillan2018,pradillo2019,sherman2020}. Under the Quincke instability, the particles spin around a random axis orthogonal to the field direction \citep{sherman2020}. When placed in an external shear flow with an electric field in the velocity-gradient direction, the rotation axis of the particles tends to align with the vorticity direction \citep{pannacci2007,dolinsky2009,das2013}. As a result, the particles spin with an angular velocity that exceeds one of the imposed flow, leading to a reduction in the effective shear viscosity. For Quincke rotors, it was shown that increasing the field strength results in a faster particle rotation and thus a further decrease in the viscosity \citep{lobry1999}.

Until now, controlling the microstructure and macroscopic rheological properties via an external electric field is mostly limited to the suspension of non-conductive particles whose responses to the electric field and shear flow have been well understood. A relatively new class of suspensions that contains conductive particles suspended in an electrolyte, which is the case of interest for this study, has gained a growing interest in timely applications, such as additive manufacturing \citep{tan2019} and electrochemical energy storage system \citep{presser2012,nikonenko2014,soloveichik2015,rommerskirchen2015,sanchez2021,folaranmi2022}. 
In such suspensions, another nonlinear electrokinetic phenomenon is expected to arise with characteristics dissimilar to one typically seen in ER fluids. Under the action of the electric field $\boldsymbol{E}$, the conductive particles can acquire an additional non-uniform charge around their surface. This charging process results in the non-uniform zeta potential distribution, which, in turn, drives a nonlinear quadrupolar flow around the particles. This flow is termed as induced-charge electroosmosis (ICEO) \citep{squires2004}. The ICEO flow is easily seen by the Helmholtz-Smoluchowski equation to scale quadratically with the magnitude of the applied electric field $E=|\boldsymbol{E}|$ because the induced zeta potential or additional surface charge is solely driven by the electric field. While the ICEO flow is perfectly symmetric for a single spherical particle, symmetric breaking can arise for a suspension due to a disturbance from other particles. The broken symmetries then result in the motion of the conductive particles, which is termed as induced-charge electrophoresis (ICEP) \citep{squires2004,squires2006}. This nonlinear electrokinetic phenomenon was first described by Murtsovkin and coworkers \citep{gamayunov1986,dukhin1986,murtsovkin1996}. Unlike the ER fluid, the particles undergoing ICEP show transient pairing dynamics, where they tend to approach along the field direction, pair up, and eventually separate in the transverse directions \citep{saintillan2008}. It is important to point out that when the suspension of conductive particles is placed in an external electric field, the particle motions arise undergoing both ICEP and DEP concurrently, which is sometimes referred to as dipolophoresis (DIP) \citep{shilov1981}. Nevertheless, it was found that DIP is mainly governed by ICEP for ideally conductive particles in a low-frequency field because of the slower decay of ICEP interactions as $O(R^{-2})$ with separation distance $R$ compared to $O(R^{-4})$ for DEP interactions \citep{saintillan2008,park2010,kilic2011}.
  
For zero-shear-limiting suspensions, we found that DIP exhibits intriguing non-trivial behaviours at the concentrated regime around a volume fraction $\phi$ in a range of $35\% \le \phi \le 50\%$ \citep{mirpark2019,mirfendereski2021}. More specifically, the hydrodynamic diffusivity and particle velocity fluctuation start to increase at $\phi \approx 35\%$ despite the expectation that they should continue to decrease with increasing $\phi$ due to the increase in the magnitude of excluded volume interactions \citep{mirpark2019}. They then reach a local maximum at $\phi \approx 45\%$ before decreasing again toward zero as $\phi$ approaches random close packing. Interestingly, such non-trivial dynamics was associated with the onset of negative particle pressure at $\phi \approx 30\%$, which eventually becomes maximum at $\phi \approx 45\%$ \citep{mirfendereski2021}. A mechanism for these counter-intuitive behaviours was explained by the transition in the dominant mechanism of particle paring arising at the concentrated regime. For a dilute DIP suspension, the particle pairing is mostly dominated by attractive particle contacts along the field direction, and contact motions are relatively slow. For semi-dilute and dense DIP suspensions, however, there are massive, very strong, and fast repulsive particle contacts in the direction perpendicular to an electric field, which is believed to be a driving mechanism for the non-trivial behaviours \citep{mirpark2019}.

While it has been focused mainly on the zero-shear-rate limit for dipolophoresis, not much effort has been made to determine the shear rheology of suspensions undergoing dipolophoresis, for which the present work aims at providing the first insight. In this paper, we develop and use large-scale numerical simulations to explore the rheology of dense suspensions of non-colloidal conductive particles undergoing both dipolophoresis and shear flow. The current simulation model is based on the Stokesian dynamics technique incorporating the mobility-based methodology for DIP \citep{saintillan2005,mirpark2019} and the resistance-based methodology for shear flow \citep{bossis1984,brady1988,durlofsky1987}. In {\S}~\ref{sec:method3}, the simulation model is described in more detail. We provide the simulation results in {\S}~\ref{sec:results3}, where we present the shear viscosity and normal stress differences of the suspension followed by suspension microstructure and dynamics. We then conclude in {\S}~\ref{sec:conclude3}.

\begin{figure}
  \centerline{
  \includegraphics[width=4.0 in]{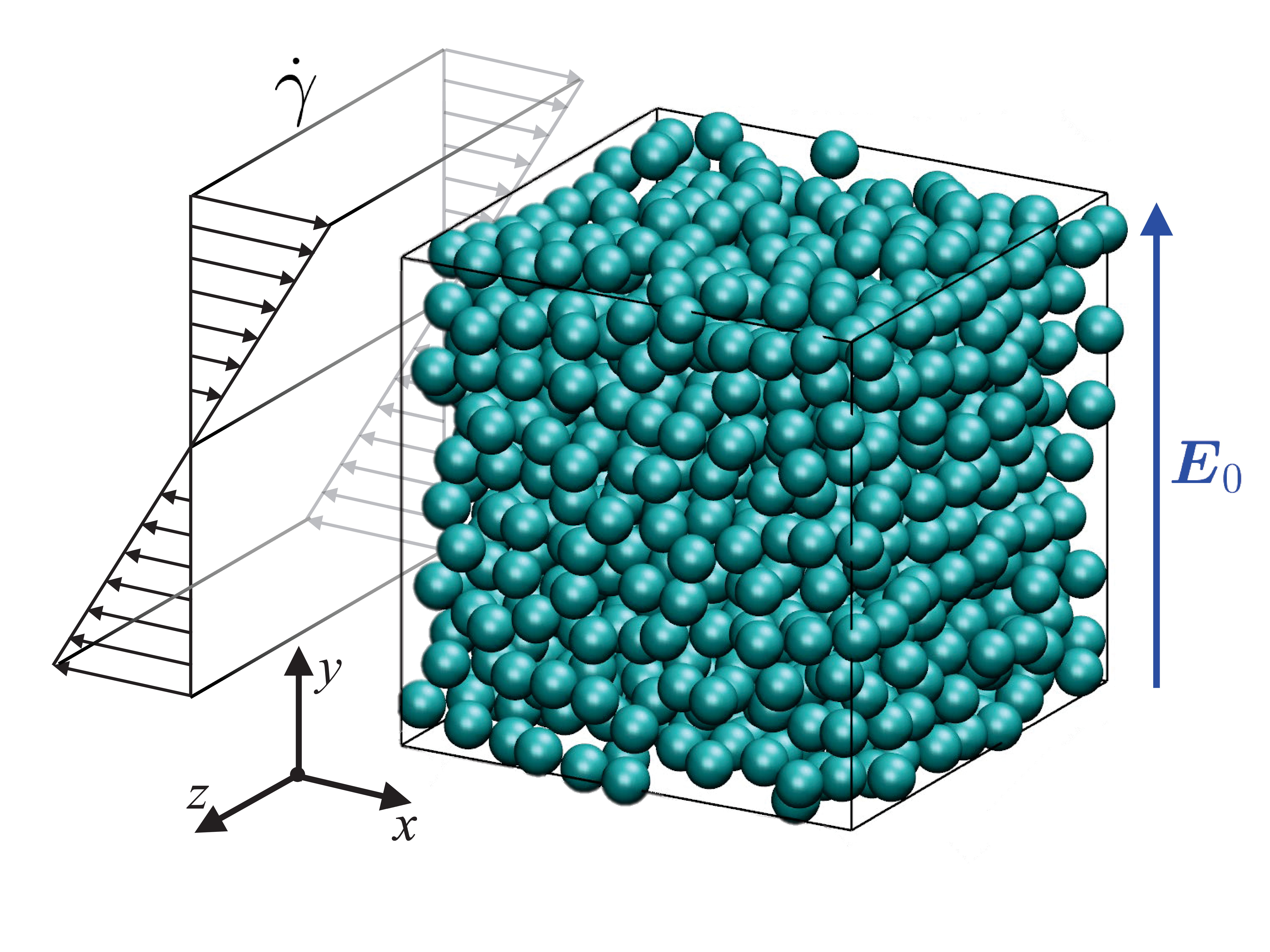}}
  \caption{A suspension of ideally conductive spheres under an external shear flow with contant shear rate $\dot{\gamma}$ and a uniform electric field $\boldsymbol{E}_0$. Note that the direction of the applied electric field is set to align with the velocity-gradient direction of the external shear flow, which is orthogonal to the flow direction.}\label{fig:fig1}
\end{figure}

\section{Governing equations and simulation method}\label{sec:method3}
We consider a suspension of $N$ identical neutrally buoyant spheres of radius $a$ in a viscous electrolyte with permittivity $\varepsilon$ and viscosity $\eta$. The spheres are considered smooth, ideally conductive or polarizable, and sufficiently large so that the Brownian motion is negligible. As demonstrated in figure~\ref{fig:fig1}, the suspension experiences a uniform electric field $\boldsymbol{E}_0 = E_0 \hat{\boldsymbol{y}}$ and simple shear flow $\boldsymbol{u}^\infty ={\mathsfbi{K}}^{\infty}\cdot \boldsymbol{x}$, simultaneously, where the $x$, $y$, and $z$ axes represent the flow, velocity-gradient, and vorticity directions of the external shear flow, respectively. Note that the electric field is along with the velocity-gradient direction and orthogonal to the flow direction. The velocity gradient tensor ${\mathsfbi{K}}^{\infty}$ of the shear flow can be given by
\begin{equation}
 {\mathsfbi{K}}^{\infty} =  
 \left[\begin{array}{ccc}  0 & \dot{\gamma} & 0 \\ 0 & 0 & 0 \\ 0 & 0 & 0 \end{array} \right],
\end{equation}
where the shear rate $\dot{\gamma}$ is constant for a steady shear flow. We use a cubic periodic domain of Lees-Edwards kind \citep{lees1972} to accommodate the bulk shear flow for the simulation of an unbounded infinite suspension.

The particles are assumed to carry no net charge, so linear electrophoresis is not expected to occur under the applied electric field. We also assume weak electric fields, thin Debye layers, and zero Dukhin number for no surface conduction \citep{squires2004}. Under these conditions and assumptions, the applied electric field drives the relative motion of the particles entirely through dipolophoresis (DIP), which is the combination of induced-charge electrophoresis (ICEP) and dielectrophoresis (DEP) \citep{saintillan2008,park2010}. For ICEP, it is important to note that the charging time of the Debye layer is very fast on the order of $\tau_c = \lambda_D a/D$, where $\lambda_D$ is the Debye layer thickness, and $D$ is the characteristic diffusivity of ions in solution. For instance, the charging time $\tau_c \sim 10^{-4}s$ in a typical experiment ($a = 10\mu$m, $\lambda_D = 10$nm, $D\sim 10^{-5}$ cm$^2$s$^{-1}$). Therefore, with the reasonable assumption of $\tau_c \ll \dot{\gamma}^{-1}$, we can easily ignore the distortion of the equilibrium charge cloud due to shear-induced rotation/translation of spheres and any consequent effects or potential instabilities \citep{khair2019}. \SM{In other words, while the particles undergo shear-induced rotation, the surface charge distribution and the screening charge cloud remain intact with respect to the external electric field.}

The current simulation model captures the contributions of both DIP and shear flow to the particle motion and suspension rheology, which are detailed as follows.

 \subsection{Electrokinetics contribution}
For computing the electric and hydrodynamic interactions resulting from both ICEP and DEP (i.e., DIP), we employ the method used in our previous studies \citep{park2010,park2011d,mirpark2019,mirfendereski2021}. Based on pair interactions calculated by \cite{saintillan2008} for a suspension undergoing DIP, the translational velocity of a given particle $\alpha$ driven by ICEP and DEP under an electric field $\boldsymbol{E}_0 = E_0 \hat{\textbf{\textit{y}}}$ can be expressed as  
\begin{equation}
\boldsymbol{U}^{d}_{\alpha} = \frac{\varepsilon a E_{0}^{2}}{\eta} \sum_{\beta=1}^{N} \left[{\mathsfbi{M}}^{DEP}(\textbf{\textit{R}}_{\alpha \beta}/a)+{\mathsfbi{M}}^{ICEP}(\textbf{\textit{R}}_{\alpha \beta}/a)\right] \boldsymbol{:}\hat{\textbf{\textit{y}}}\hat{\textbf{\textit{y}}}, \ \ \ \ \ \alpha=1,...,N,\label{eq:sums3}
 \end{equation} 
where $\textbf{\textit{R}}_{\alpha \beta} = \textbf{\textit{x}}_{\beta} - \textbf{\textit{x}}_{\alpha}$ is the separation vector between the particle $\alpha$ and particle $\beta$, and ${\mathsfbi{M}}^{DEP}$ and ${\mathsfbi{M}}^{ICEP}$ are third-order dimensionless tensors accounting for the DEP and ICEP interactions, respectively. It is shown that these two tensors are entirely determined by the scalar functions of the dimensionless inverse separation distance $\lambda= 2a/\vert \textbf{\textit{R}} \vert$. For far-field interactions ($\lambda \ll 1$), the DEP and ICEP far-field interaction tensors can be computed using the method of reflections and expressed up to order $ O(\lambda^4)$ in terms of two fundamental solutions of Stokes equations ${\boldsymbol{\mathsf{S}}}$ and ${\boldsymbol{\mathsf{T}}}=\nabla^{2}{\boldsymbol{\mathsf{S}}}$, which are the Green's functions for a Stokes dipole and for a potential quadrupole, respectively \citep{kim2013}. The far-field expressions of these tenors are written as follows:
\begin{equation}
{\mathsfbi{M}}_{FF}^{DEP} = \frac{1}{12}{\boldsymbol{\mathsf{T}}} + O(\lambda^5),
\label{eq:dep}
\end{equation}
\begin{equation}
{\mathsfbi{M}}_{FF}^{ICEP} = -\frac{9}{8}{\boldsymbol{\mathsf{S}}}-\frac{11}{24}{\boldsymbol{\mathsf{T}}} + O(\lambda^5).
\label{eq:icep}
\end{equation}
These tensors can be expressed as the periodic version of the far-field tensors to account for far-field interactions between particles $\alpha$ and $\beta$ and all its periodic images, which are valid to order $O(\boldsymbol{R}_{\alpha \beta}^{-4})$. Direct calculation of the sums for $\boldsymbol{U}^{d}_{\alpha}$ in (\ref{eq:sums3}) requires $O(N^2)$ of computation for which the smooth particle mesh Ewald algorithm, which shares similarities to the accelerated Stokesian dynamics simulation \citep{Sierou2001}, is used to accelerate the calculation of the sums to $O(N\textrm{log}N)$ operations \citep{mirpark2019}. However, the near-field corrections are necessary as the method of reflections becomes inaccurate when the particles are close to each other (typically for $|\textbf{\textit{R}}_{\alpha\beta}| < 4a$). This is achieved by correcting the far-field tensor with the more accurate method of twin multiple expansions \citep{saintillan2008}. This method is very accurate down to separation distances on the order of $ \left | \boldsymbol{R}_{\alpha \beta} \right | \approx 2.005a$.

\SM{For DEP and ICEP interactions due to (\ref{eq:dep}) and (\ref{eq:icep}), it is worth noting that while both ICEP and DEP lead to similar particle pairing in the electric-field direction, the subsequent paring dynamics is significantly different \citep{saintillan2008,park2010}. Specifically, ICEP leads to transient pairing dynamics, where particles briefly pair up in the field direction but then reorient themselves to align in the transverse directions, eventually leading to their separation. Such paring dynamics give rise to chaotic collective motion. Conversely, in the DEP case, the formed pairs remain in a stable equilibrium, ultimately resulting in the formation of stable chains/columns along the field direction. In the case of DIP, when both ICEP and DEP coexist, transient pairing dynamics only occur due to the predominance of ICEP over DEP, as can be alluded by the leading orders of the ICEP and DEP interactions, which are a Stokes dipole of $O(\lambda^2)$ and a potential quadrupole of $O(\lambda^4)$, respectively.}

\subsection{Shear flow contribution}
The contribution of the external shear flow $\boldsymbol{u}^{\infty}$ to the particle motion and bulk stress is calculated by the classical resistance-based Stokesian dynamics approach \citep{brady1988}. 
Owing to the linearity of the Stokes equation, we can relate the moments of the hydrodynamic force density on the particle surfaces to the relative velocity moments of particles by the grand resistance tensor $\mathcal{R}$ as given by 
\begin{equation}
 \left[\begin{array}{ccc}  \boldsymbol{F} \\ \boldsymbol{T} \\ {\mathsfbi{S}} \end{array} \right] = -
 \mathcal{R}
  \boldsymbol{\cdot} \left[\begin{array}{ccc}  \boldsymbol{U}-\boldsymbol{u}^{\infty} \\ \boldsymbol{\Omega}- \boldsymbol{\omega}^{\infty} \\ -{\mathsfbi{E}}^{\infty} \end{array} \right],
\end{equation}
where $\boldsymbol{U}$ and $\boldsymbol{\Omega}$ are translational and rotational velocities of particles, respectively, and $\boldsymbol{F}$, $\boldsymbol{T}$, and $\mathsfbi{S}$ are the force, torque, and stresslet on particles, respectively. The rate of strain ${\mathsfbi{E}}^{\infty}$ is the symmetric part of ${\mathsfbi{K}}^{\infty}$, and $\boldsymbol{\omega}^{\infty} = 1/2 \boldsymbol{\nabla \times u}^{\infty}$ gives the vorticity of the imposed flow. The grand resistance tensor $\mathcal{R}$ can be approximated as a sum of the Stokes drag and the pairwise lubrication interactions since the dominant hydrodynamic interactions come from the pairwise short-range lubrication forces for dense suspensions \citep{mari2014}. The grand resistance tensor $\mathcal{R}$ can be then written by the combination of the resistance matrices \citep{bossis1987}:
\begin{equation}
\mathcal{R} = \left[\begin{array}{ccc}  {\mathsfbi{R}}_{FU} & {\mathsfbi{R}}_{FE} \\  {\mathsfbi{R}}_{SU} & {\mathsfbi{R}}_{SE}  \end{array} \right],
\end{equation}
where the $6N\times6N$ second-rank tensor ${\mathsfbi{R}}_{FU}$ relates the hydrodynamic forces/torques on particles to their motion (translation/rotational velocity) relative to the imposed flow. The third-rank tensors ${\mathsfbi{R}}_{FE}$ and ${\mathsfbi{R}}_{SU}$ relate the hydrodynamic forces/torques to the rate of strain and the particle stresslet to the relative motion of particles, respectively. The fourth-rank tensor ${\mathsfbi{R}}_{SE}$ gives the particle stresslet owing to the rate of strain. These resistance tensors are configuration-dependent, and their elements can be expressed by the relationships between the unit vector along the center line of each pair of particles $\hat{\boldsymbol{R}} = \boldsymbol{R}_{\alpha \beta}/|\boldsymbol{R}_{\alpha \beta}|$ and the scalar resistance functions of $\lambda$. The details of calculating these resistance tensors were provided by \cite{jeffrey1984, kim1985,jeffrey1992, kim2013}. 

\subsection{Total particle velocity}
We calculate the total particle velocity and stress by superposing the contributions of DIP and shear flow thanks to the linearity of the Stokes equation. Hence, the total translational velocity $\boldsymbol{U}^t$ and rotational velocity $\boldsymbol{\Omega}^t$ of the particles are given by 
\begin{equation}\label{eq:vel3}
\left[ \begin{array}{ccc} \boldsymbol{U}^t \\ \boldsymbol{\Omega}^t \end{array} \right] = 
\left[ \begin{array}{ccc} \boldsymbol{u^{\infty}} \\ \boldsymbol{\omega}^{\infty} \end{array} \right] + {\mathsfbi{R}}^{-1}_{FU}\boldsymbol{\cdot} {\mathsfbi{R}}_{FE}\boldsymbol{:}{\mathsfbi{E}}^{\infty} + \mathsfbi{R}^{-1}_{FU} \boldsymbol{\cdot} \boldsymbol{F}^p + \left[ \begin{array}{cc} \boldsymbol{U}^d \\ \boldsymbol{\Omega}^d \end{array} \right],
\end{equation}
where $\boldsymbol{\Omega}^d = \boldsymbol{0}$ as no particle rotation arises as a result of DIP for the perfectly symmetric spheres \citep{saintillan2008}. The explicit calculation of the second and third terms in the right-hand side of (\ref{eq:vel3}) is prohibitively expensive due to an inversion of the resistance tensor ${\mathsfbi{R}}_{FU}$, which is of an order of $O(N^3)$. For the sake of fast calculations, we use the generalized minimal residual (GMRES) method \citep{saad1986} to resolve these terms. The interparticle force $\boldsymbol{F}^p$ in (\ref{eq:vel3}) is written as a pairwise electrostatic repulsive force between the particles $\alpha$ and $\beta$ \citep{bossis1984,sierou2002}:
\begin{equation}\label{eq:interforce}
 \boldsymbol{F}_{\alpha \beta}^{p} = \frac{F_0}{\sigma} \frac{ e^{-\epsilon/\sigma }}{1-e^{-\epsilon/\sigma }} \hat{\boldsymbol{R}}_{\alpha \beta},
\end{equation}
where $F_0$, $\sigma$, and $\epsilon$ are the force magnitude, force range, and the dimensionless spacing between the surface of particles $\epsilon = R/a - 2$, respectively. The magnitude of the interparticle force relative to the shear force can be controlled by the dimensionless number $\dot{\gamma}^* = 6 \pi \eta a^2 \dot{\gamma}/F_0$. For all present simulations, the fixed values of $\dot{\gamma}^* = 5,000$ and $\sigma = 0.001$ are employed as validated and used by \cite{sierou2002}. \SM{It is important to note that the choice of these values for the repulsive force corresponds to a very short time scale with a negligible impact for the system studied.}

Once the particle velocities are calculated by (\ref{eq:vel3}), the particle positions are advanced in time using a second-order Adams-Bashforth time-marching scheme, with an explicit Euler scheme for the first time step. A fixed time step $\Delta t$ is chosen so as to ensure that the particles only travel a very small fraction ($<0.4\%$) of the mean interparticle distance at each iteration. However, due to the use of finite time step, the particle overlap should be prevented, for which we employ the potential-free algorithm \citep{melrose1993} from our previous works \citep{mirpark2019,mirfendereski2021}. \SM{The initial random equilibrium configurations were generated using an approach similar to the one employed in our previous works \citep{mirpark2019,mirfendereski2021}. This method is based on the procedure suggested for dense hard-sphere systems \citep{stillinger1964,clarke1987,rintoul1996}.} 

\subsection{Bulk stress}
The bulk rheological properties can be determined from the average stress tensor $\left\langle {\mathsfbi{\Sigma}} \right\rangle$ \citep{batchelor1970}, where the angle bracket denotes the ensemble average over all particles. The stress tensor is expressed as
\begin{equation}\label{eq:bulkstress3}
\left \langle \mathsfbi{\Sigma} \right \rangle = -p_f{\mathsfbi{I}} +2\eta{\mathsfbi{E}}^{\infty}+\left \langle \mathsfbi{\Sigma}^p \right \rangle,
\end{equation}
where $p_f$ is the \SM{is the average pressure in the fluid phase}, ${\mathsfbi{I}}$ is the identity matrix, and $2\eta{\mathsfbi{E}}^{\infty}$ is the deviatoric contribution from the incompressible fluid. The total contribution of particle phase to the stress or particle stress $\left\langle{\mathsfbi{\Sigma}}^p\right\rangle$ is given by
 \begin{equation}\label{eq:stress3}
 \left\langle\mathbf{\Sigma}^p\right\rangle = -n\left\langle {\mathsfbi{R}}_{SU} \boldsymbol{\cdot} \left[\begin{array}{cc} \boldsymbol{U}-\boldsymbol{u}^{\infty} \\  \boldsymbol{\Omega}-\boldsymbol{\omega}^{\infty} \end{array}\right] \right\rangle + n\left\langle {\mathsfbi{R}}_{SE}\boldsymbol{:}{\mathsfbi{E}}^{\infty} \right\rangle + n\left \langle {\mathsfbi{S}}^{d} \right\rangle - n\left\langle \boldsymbol{x} \boldsymbol{F}^p \right\rangle,
 \end{equation}
where $n$ is the particle number density. The first and second terms in (\ref{eq:stress3}) are the hydrodynamic stresslets that come from the external shear flow \citep{sierou2002,mari2014}, while the third term $\left \langle {\mathsfbi{S}}^{d} \right\rangle$ stems from the DIP contribution \citep{mirfendereski2021}. Note that the contribution of DEP to the stress is negligible compared to the ICEP contribution for ideally conductive particles \citep{mirfendereski2021}. The direct interparticle force contribution is given by $-\langle \boldsymbol{xF}^p\rangle$ in (\ref{eq:stress3}), which is found to be negligible here.

\begin{figure}
  \centerline{
  \includegraphics[width=5.2 in]{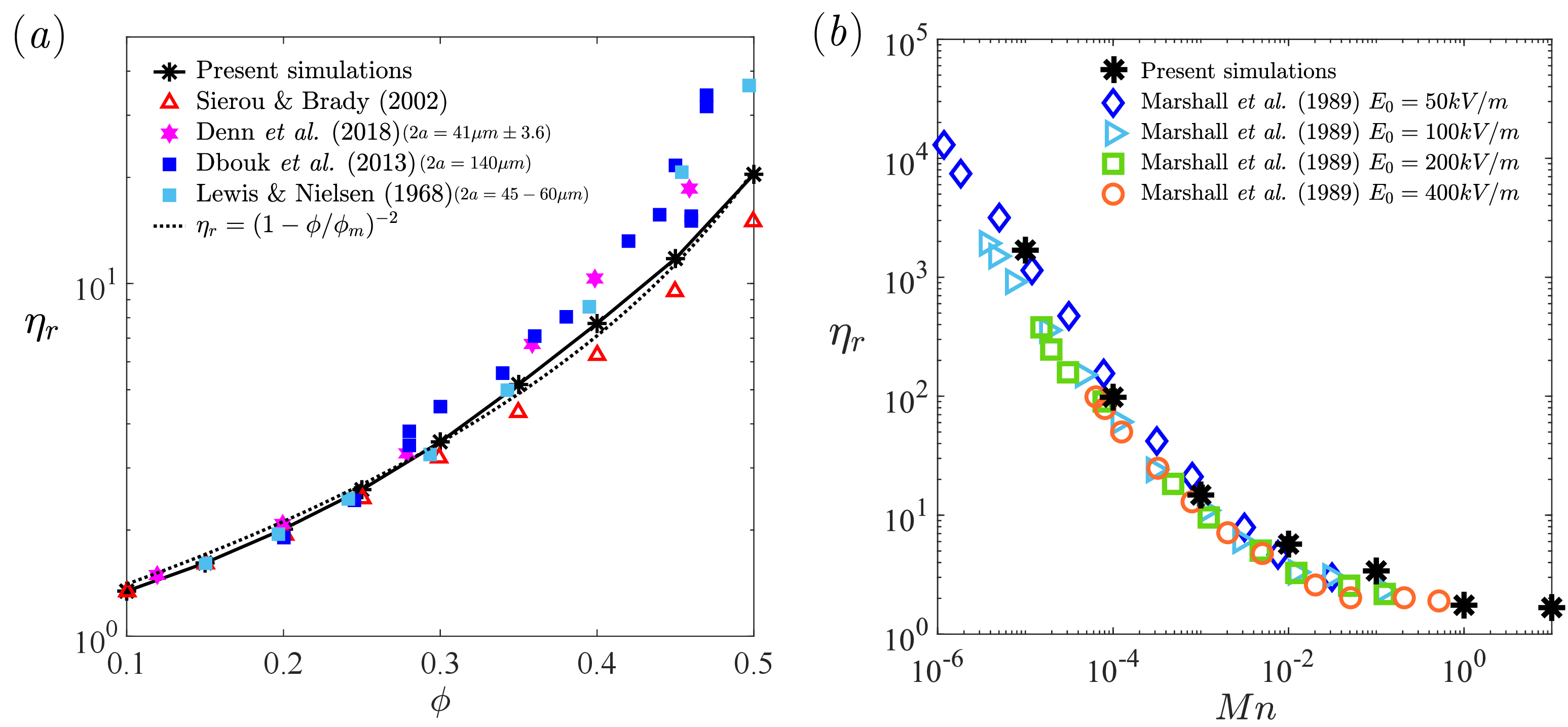}}
  \caption{\SM{(\textit{a})} The relative shear viscosity $\eta_r$ of non-Brownian suspension solely undergoing shear flow as a function of volume fraction. Also shown are the experimental results of \cite{lewis1968}, \cite{dbouk2013}, and \cite{denn2018} and numerical results of \cite{sierou2002}. The current results agree well with the numerical and experimental data reported in the literature. The dotted line is the Maron-Pierce correlation with $\phi_m = 0.63$. \SM{(\textit{b}) Relative shear viscosity of the suspension undergoing DEP and shear flow as a function of Mn at $\phi = 15\%$ from the current simulations. Also shown are the experimental results of \cite{marshall1989} for electrorheological (ER) suspensions containing hydrated poly(methacrylate) particles in a chlorinated hydrocarbon subjected to different electric field strengths at $\phi = 13\%$.}}\label{fig:fig2}
\end{figure}

Once the average bulk stress tensor is obtained, its $xy$ component  $\langle \Sigma_{xy} \rangle$ is used to calculate the relative shear viscosity:
\begin{equation}
\eta_r  = \frac{\langle \Sigma_{xy} \rangle}{2\eta E^{\infty}_{xy}}.
\end{equation}
The normal components of the average bulk stress are also used to calculate the first and second normal stress differences:
\begin{equation}\label{eq:N2}
N_1 = \langle \Sigma_{xx}\rangle - \langle \Sigma_{yy}\rangle,
\end{equation}
\begin{equation}\label{eq:N1}
N_2 = \langle \Sigma_{yy}\rangle - \langle \Sigma_{zz}\rangle. 
\end{equation}
For testing the code for validation purposes, the relative shear viscosity in the absence of an electric field is calculated using the present simulation model, which is shown in figure~\ref{fig:fig2}\SM{(\textit{a})}. In addition to the current simulation results, the experimental results by \cite{dbouk2013,denn2018,lewis1968} and computational results by \cite{sierou2002} are present along with the Maron-Pierce correlation $\eta_r = (1- \phi / \phi_m)^{-2}$ with $\phi_m = 63\% $. A good agreement between the current and previous results is observed. In particular, the current simulation model reproduces the Maron-Pierce correlation well. 

To indicate the relative importance of the viscous stress from shear flow over the electrokinetics-driven stress from DIP, the so-called Mason number ($Mn$) can be used and given by
\begin{equation}
Mn=\frac{\eta \dot{\gamma}}{\varepsilon E_0^2},
\end{equation}
where the viscous shear stress and DIP stress scale as $\eta \dot{\gamma}$ and $\varepsilon E_0^2$, respectively.

\SM{To further validate the simulation code in the presence of both electric field and shear flow, we conducted simulations for suspensions undergoing shear flow and DEP, where ICEP is absent, a scenario closely resembling typical electrorheological (ER) fluids. Figure~\ref{fig:fig2}(\textit{b}) illustrates the relative shear viscosity as a function of the Mason number, presenting the current simulation results along with the experimental data reported by \cite{marshall1989} for ER fluids. A good agreement between the current computational results and the previous experimental data is observed.}

\section{Results and discussion}\label{sec:results3}
We performed large-scale simulations of ideally conductive particle suspensions under both electric field and shear flow in a periodic cubic domain of Lees-Edwards kind in ranges of volume fractions and Mason numbers. \SM{The dependence on the system size was examined, confirming that the simulation results are insensitive to the domain size and the number of particles.} It is worth noting again that the Mason number is the ratio of viscous shear stress and electrokinetics-driven stress, representing the dimensionless shear rate for the system studied here. \SM{However, it can also be regarded that given the shear rate, the Mason number represents the inverse of the squared dimensionless field strength.} All simulation runs were initiated from the hard-sphere equilibrium configurations, and the time steps within the first 10--30 strains were discarded for statistical averaging over the steady-state configurations. \SM{To reach 200 strains $\gamma$, the current simulation at $\phi = 15\%$ requires approximately 50 hours of the CPU time, and a higher volume fraction at $\phi = 50\%$ requires approximately 7 times more CPU times as compared to  $\phi = 15\%$. In addition, it is worth noting that the CPU time increased with higher volume fractions and lower Mason numbers.}  The statistical variation of the suspension properties over the steady state is determined by their standard deviation and represented by the error bar in the figures. 

\begin{figure}
  \centerline{
  \includegraphics[width=5.3 in]{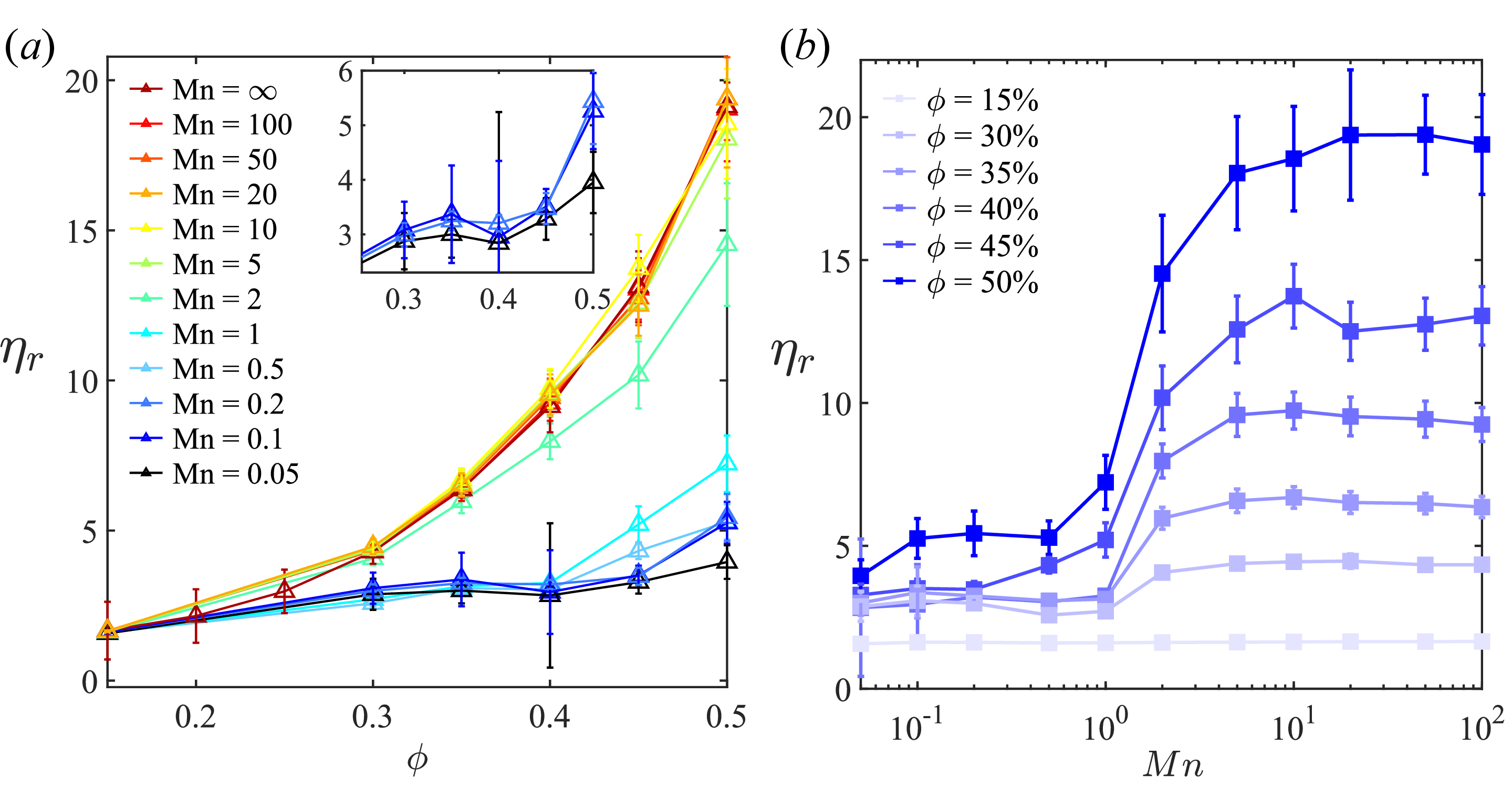}}
  \caption{(\textit{a}) The relative shear viscosity $\eta_r$ of the non-Brownian suspensions undergoing both dipolophoresis and steady shear flow as a function of volume fraction for a range of Mason numbers. Note that $Mn = \infty$ limit refers to the viscosity in the absence of an electric field. Inset: the magnified view of the low-$Mn$ or low-shear-rate viscosities ($Mn = 0.05, 0.1, 0.2$), showing a non-monotonic variation with volume fraction. (\textit{b}) The relative shear viscosity is replotted as a function of Mason number for various volume fractions.}\label{fig:fig3}
\end{figure}

\subsection{Relative shear viscosity}
We start by exploring the effects of the Mason number and volume fraction on the shear viscosity of suspensions. Figure~\ref{fig:fig3}(\textit{a}) presents the relative shear viscosity $\eta_r$ as a function of volume fraction for a range of Mason numbers. As seen in the figure, the high-shear-rate viscosities ($Mn\geqslant5$) collapse almost onto a single curve. Note that $Mn = \infty$ refers to a measurement in the absence of an electric field. By reducing the shear rate or the Mason number below $Mn = 5$, the viscosity starts to decrease, more significantly at high volume fractions for $\phi \geqslant 40\%$. Interestingly, there is a significant decrease from $Mn = 2$ to $Mn = 1$, suggesting a possible transition taking place around these Mason numbers. At the low-shear-rate regime ($Mn < 0.5$), the viscosity seems to collapse again onto a single curve with smaller values than for the high-shear-rate regime. The inset of the figure shows a magnified view of the low-shear-rate viscosities ($Mn = 0.2, 0.1, 0.05$) as a function of volume fraction. Interestingly, a non-monotonic variation with volume fraction is observed. These low-shear-rate viscosities start to increase with volume fraction, reach a local maximum at $\phi \approx 35\%$, and then slightly decrease up to $\phi \approx 40\%$ before increasing again as volume fraction is further increased. Such non-monotonic behaviour is attributed to the predominant ICEP effect over shear flow at low Mason numbers or low shear rates. This non-monotonic trend can be explained by quiescent DIP suspensions in our previous studies, where the hydrodynamic diffusivity exhibits the exact opposite trend to the shear viscosity \citep{mirpark2019} and the particle pressure exhibits a very similar non-monotonic trend \citep{mirfendereski2021}.
 


To characterize a shear-dependence behaviour, the relative shear viscosity is replotted as a function of the Mason number for various volume fractions, as seen in figure~\ref{fig:fig3}(\textit{b}). At a volume fraction as small as $\phi = 15\%$, the variation of viscosity with Mason number is almost negligible, indicating a Newtonian-fluid-like behaviour. For $\phi \geqslant 30\%$, however, the shear-thickening behaviour is observed. The viscosity appears to remain constant at low Mason numbers and starts to sharply increase around $Mn = 1$, reaching a plateau as the Mason number is further increased. Similar to figure~\ref{fig:fig3}(\textit{a}), a significant transition again seems to occur at $Mn \approx 1$, which can be thus referred to as the transitional or critical Mason number for the current study. The shear thickening becomes more prominent with increasing volume fraction. At $\phi = 45\% - 50\%$, a large increase in viscosity occurs during the shear-thickening process, which is about five-fold. The underlying mechanism of such shear thickening in the concentrated regime is intimately linked to the emergence of distinct microstructure, for which we will discuss in {\S}~\ref{subsec:micro}. As an interesting future work, this $Mn$-dependent behaviour suggests a possible rheology control for a suspension of ideally conductive particles in an electric field. For example, at a given high volume fraction (i.e., $\phi \ge 30\%$) and high shear rate (i.e., $Mn > 1$), having stronger DIP or ICEP effects via increasing the electric-field strength could lead to a viscosity reduction from the shear-thickened states to the Newtonian-like low viscosity states. 

It is interesting to note that the shear-thickening behaviour of the current DIP suspension is in contrast to what has been typically observed for the classical electrorheological (ER) fluid, which mainly exhibits shear-thinning behaviour \citep{bonnecaze1992dynamic,parthasarathy1996}. This distinct difference in shear-dependent rheological responses should be associated with the predominance of ICEP over DEP in the DIP suspension of ideally conductive particles \citep{park2010,park2011d}. Note that the underlying mechanism of the ER fluid ties closely to DEP interactions \citep{mirfendereski2022}. 

\begin{figure}
  \centerline{
  \includegraphics[width=5.4 in]{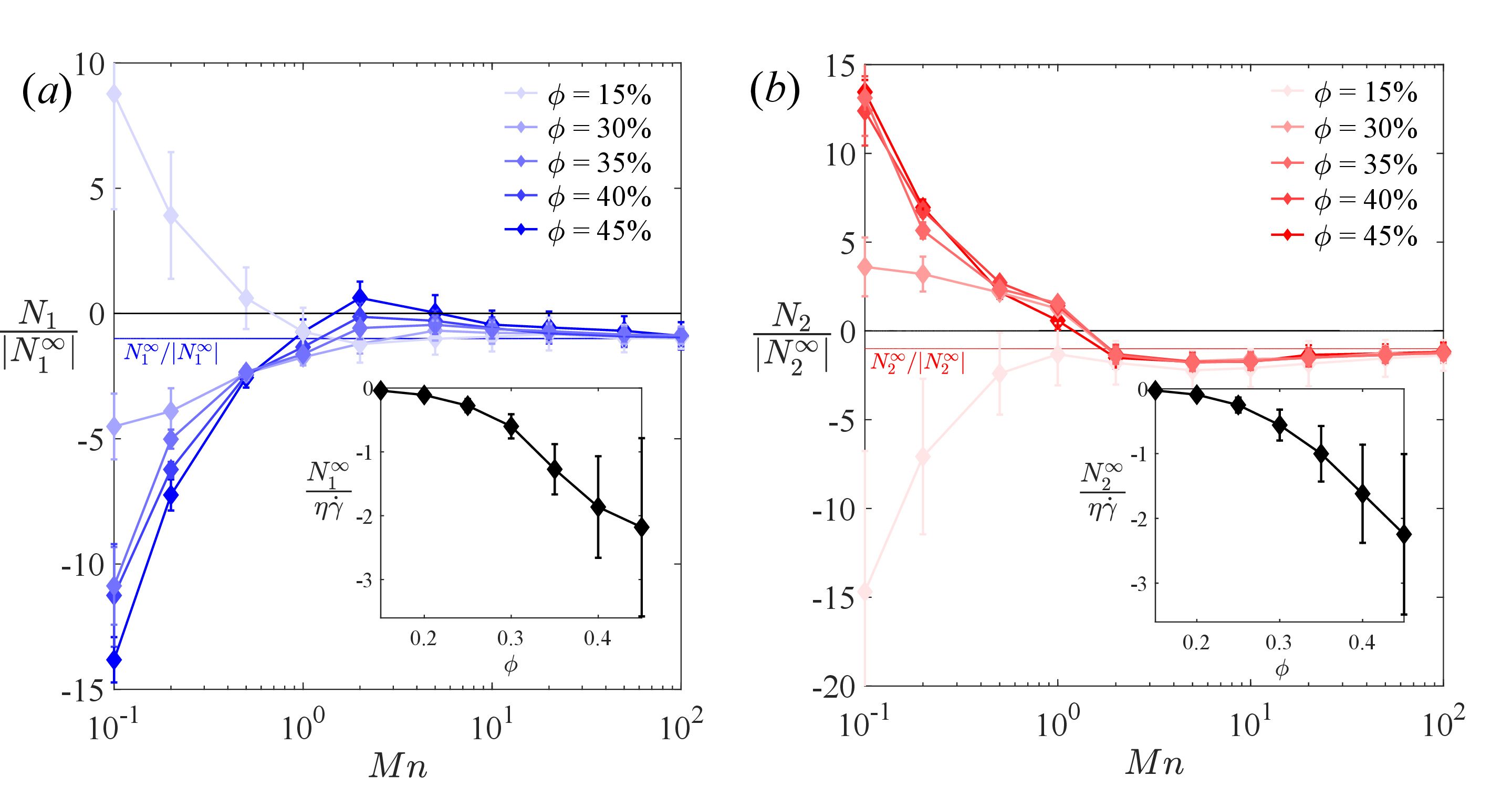}}
  \caption{The (\textit{a}) first and (\textit{b}) second normal stress differences, normalized by their magnitude for the pure-shear-flow limit ($Mn = \infty$), as a function of Mason number for various volume fractions. The inset of (\textit{a}) and (\textit{b}) shows the first and second normal stress differences for $Mn={\infty}$ as a function of volume fraction, respectively.}\label{fig:fig4}
\end{figure}

\subsection{Normal stress differences}
As it has been observed that significant particle normal stresses are generated by DIP even at zero shear rate \citep{mirfendereski2021}, we investigate the normal stress differences as a result of shear flow. The first and second normal stress differences, which are defined by $N_1 = \Sigma_{xx}-\Sigma_{yy}$ and $N_2 = \Sigma_{yy}-\Sigma_{zz}$, respectively \citep{zarraga2000,sierou2002,dbouk2013}, are examined. Prior to proceeding to a detailed analysis for DIP suspensions with shear flow, the insets of  figures \ref{fig:fig4}(\textit{a}) and (\textit{b}) present the pure-shear-flow limiting normal stress differences $N^{\infty}_1$ and $N^{\infty}_2$ as a function of volume fraction, respectively. As clearly seen, both $N^{\infty}_1$ and $N^{\infty}_2$ remain negative, and their magnitude increases with volume fraction, which has also been reported both experimentally and numerically for sheared, non-colloidal suspensions \citep{sierou2002,dai2013,pan2015,guazzelli2018}. It is worth noting that $N^{\infty}_1$ and $N^{\infty}_2$ at $\phi = 45\%$ are almost the same as those for high-shear colloidal dispersions \citep{foss2000} and high-shear non-colloidal dispersions \citep{sierou2002}. The notable error bars are observed at high volume fractions. These errors are attributed to the sensitivity of the normal stress differences to even small errors in the calculation of the average value of each normal stress component, which is also seen and explained by \cite{sierou2002}.

Figures~\ref{fig:fig4}(\textit{a}) and (\textit{b}) show $N_1$ and $N_2$ as a function of Mason number for various volume fractions, respectively, where the normal stress differences are normalized by their values for a pure shear flow ($Mn = \infty$). At low Mason numbers or specifically below the critical Mason number ($Mn = 1$), the DIP effect is expected to be strong relative to the shear flow. In this DIP-dominant regime, $N_1$ is positive for a volume fraction as small as $\phi = 15\%$ but negative at $\phi \geqslant 30\%$. As opposed to the first normal stress difference, $N_2$ is negative at $\phi = 15\%$ but positive at $\phi \geqslant 30\%$. The positive $N_1$ and negative $N_2$ at a low volume fraction of $\phi = 15\%$ indeed share similarities with the Brownian contributions to the normal stress differences observed in a hard-sphere colloidal dispersion \citep{foss2000}. The distinctive characteristics of the low-shear-rate $N_1$ and $N_2$ can also be explained by our previous observation of the DIP-driven normal stresses at zero shear rate \citep{mirfendereski2021}. We observed that under dipolophoresis only, the magnitude of the normal stress in the field direction is larger than the transverse ones (i.e., $|\Sigma_{yy}| > |\Sigma_{xx}|, |\Sigma_{zz}|$) for all volume fractions. In addition, the signs of field and transverse normal stresses are opposite at different regimes of volume fraction; specifically, the suspension at $\phi < 30\%$ exhibits positive transverse normal stresses ($\Sigma_{xx},\Sigma_{zz}$) and a negative field normal stress ($\Sigma_{yy}$), while the former becomes negative and the latter becomes positive at $30\% \le \phi \le 50\%$. That is again consistent with the unique low-shear-rate characteristics of $N_1$ and $N_2$ described above and shown in figure~\ref{fig:fig4}. For $Mn \gg 1$ or well above the critical Mason number, both $N_1$ and $N_2$ become negative for all volume fractions considered, which indicates that the effect of shear flow becomes dominant over the DIP effect. Both $N_1$ and $N_2$ appear to reach a high-shear plateau for $Mn > 10$. It should be noted that the negative sign of both normal stress differences at high Mason numbers is well observed in concentrated suspensions, especially in the shear-thickened states \citep{cwalina2014}.

\begin{figure}
  \centerline{
  \includegraphics[width=5.5 in]{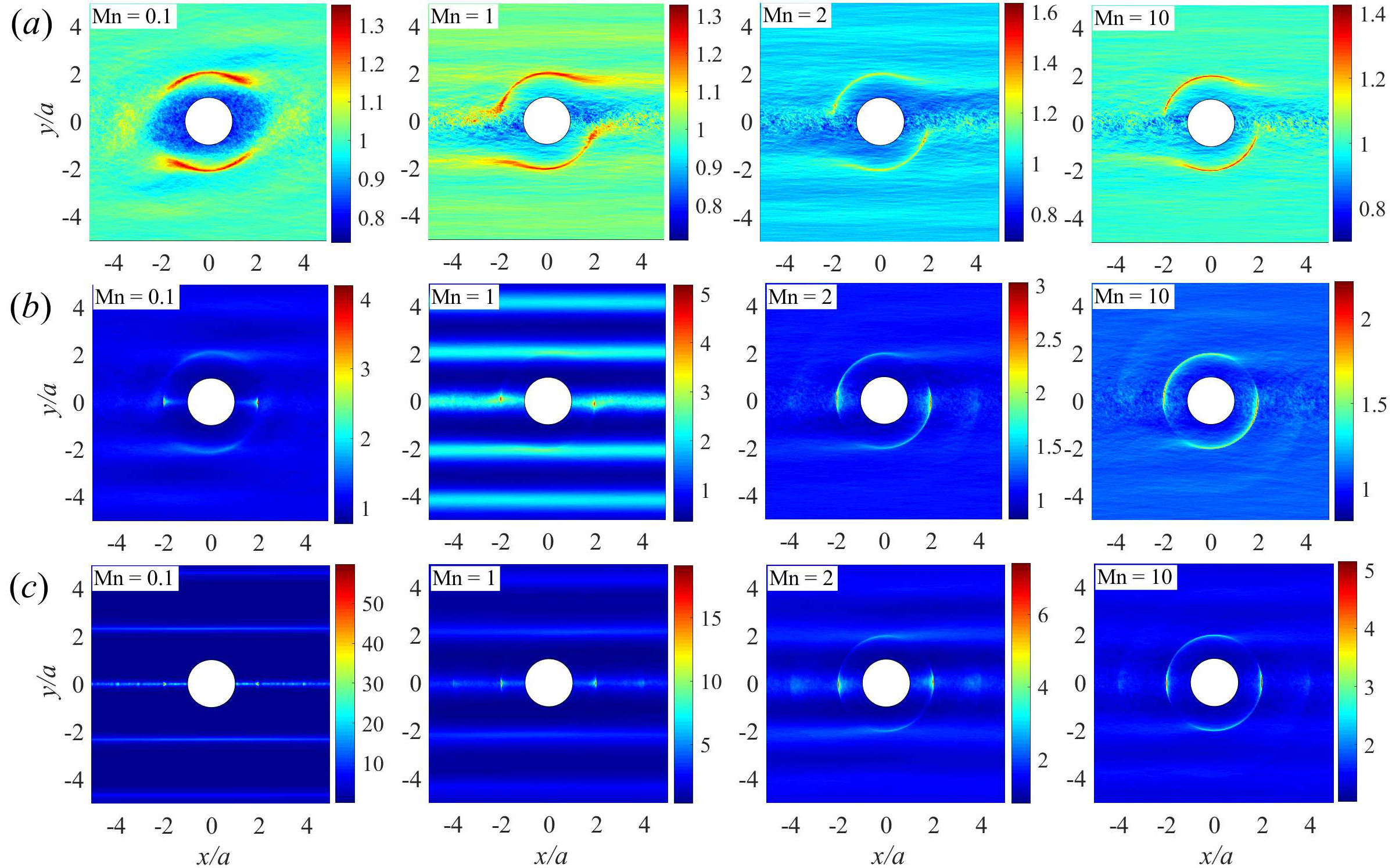}}
  \caption{The projection of the pair distribution function onto the shear plane ($x$-$y$) at various Mason numbers: (\textit{a}) top-row panel for $\phi = 15\%$, (\textit{b}) middle-row panel for $\phi = 30\%$, and (\textit{c}) bottom-row panel for $\phi = 45\%$.}\label{fig:fig5}
\end{figure}

\subsection{Suspension microstructure}\label{subsec:micro}
As the suspension microstructure has been regarded as a microscopic basis of the shear viscosity and the normal stress differences \citep{wagner1992,brady1997,zarraga2000}, we attempt to investigate the microstructural variation in connection to the variation of the macroscopic rheological properties. To directly characterize the microstructure undergoing both DIP and shear flow, we calculate the pair distribution function \citep{park2010,mirpark2019}. This function provides the probability of finding the particles with respect to a reference particle placed at the origin. Figure~\ref{fig:fig5} shows the projection of this function on the shear plane ($x$-$y$) for various Mason numbers at volume fractions of $\phi = 15\%, 30\%$, and $45\%$. At $Mn = 0.1$, which is below the critical Mason number, the function is symmetric with respect to the $y$ axis (the electric-field direction) for all volume fractions considered. However, it is interesting to note that as the volume fraction increases, the location of the maximum probability changes from near the particle poles ($\phi = 15\%$) to the particle equators ($\phi = 30\%, 45\%$), which is similar to the zero-shear-rate limiting suspension \citep{mirpark2019,mirfendereski2021}. On the relation of the microstructure to the normal stress differences, the high probability near the poles formed at $\phi = 15\%$ and $Mn = 0.1$ suggests the strong negative normal stress in the $y$ direction compared to the other two directions. This leads to the positive $N_1$ and negative $N_2$ at low Mason numbers ($Mn < 1$), as seen in figures~\ref{fig:fig4}($a$) and ($b$). On the other hand, for higher volume fractions of $\phi \geqslant 30\%$, the high probability at the equators corresponds to the strong negative normal stress in the $x$ and $z$ directions, leading to the negative $N_1$ and positive $N_2$ at low Mason numbers ($Mn < 1$), as also seen in figures~\ref{fig:fig4}($a$) and ($b$).
 
Increasing the shear rate then results in the notable distortion of the microstructure, as seen in figure~\ref{fig:fig5} for $Mn = 1, 2,$ and $10$. At a low volume fraction of $\phi = 15\%$ in figure~\ref{fig:fig5}(\textit{a}), the microstructure evolves like a typical shear-driven suspension \citep{foss2000}, where the compressive and extensional axes are clearly identified. The anisotropy in the microstructure is further developed by increasing the Mason number. This microstructural anisotropy -- the high probability along the compressional axis and the low probability along the extensional axis -- indeed results in negative normal stress differences at high shear rates \citep{foss2000,sierou2002,cwalina2014}. For a higher volume fraction of $\phi = 30\%$ in figure~\ref{fig:fig5}(\textit{b}), an intriguing microstructural feature is observed at $Mn = 1$, where a striped pattern or a string-like phase is horizontally formed in the direction perpendicular to the field ($y$) direction. This observation suggests that the particles tend to assemble into two-dimensional ($x$-$z$) structures parallel to one another with a certain distance apart. This seemingly ordered microstructure could be responsible for a slight reduction in viscosity from $Mn = 0.1$ to $Mn = 1$ in figure~\ref{fig:fig3}(\textit{b}), which could resemble shear thinning for hard-sphere colloidal suspensions \citep{phung1996,wagner2009}. For a much higher volume fraction of $\phi = 45\%$ in figure~\ref{fig:fig5}(\textit{c}), the horizontal striped patterns also emerge but form earlier at $Mn < 1$ than for $\phi = 30\%$. A further increase in shear rate again leads to the microstructure being analogous to the typical shear-driven suspension for both $\phi = 30\%$ and $45\%$, showing the anisotropic microstructure with respect to the compressive and extensional axes. Interestingly, the ordered patterns start to break around the critical Mason number ($Mn = 1$), after which so-called hydroclusters are formed. A further rise in the viscosity during the shear-thickening process at $\phi = 45\%$ compared to $\phi = 30\%$ could be attributed to denser hydroclusters, which can be elucidated by the pair distribution functions at $Mn = 10$ in figures~\ref{fig:fig5}($b$) and ($c$). Accompanying supplementary movies show the particle motions in a suspension for $Mn = 0.1, 1,$ and $10$ at $\phi = 30\%$ and $45\%$.

\begin{figure}
  \centerline{
  \includegraphics[width=3.0 in]{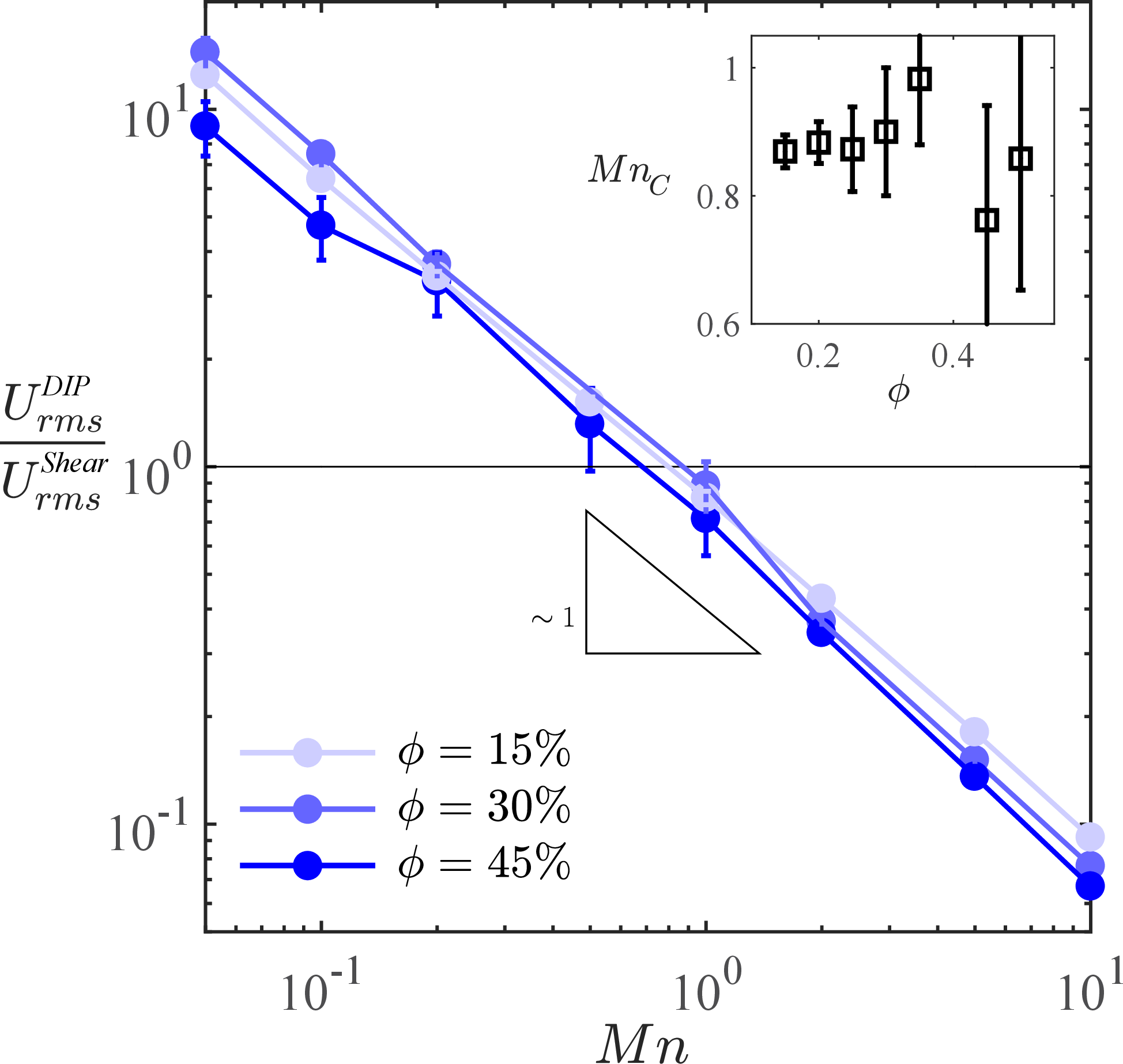}}
  \caption{The ratio of the dipolophoresis contribution and the shear contribution to the root-mean-square (RMS) velocity of particles ($U^{DIP}_{rms}/U^{Shear}_{rms}$) as a function of Mason number on a log-log scale. Inset: the crossover Mason number $Mn_c$ as a function of volume fraction. Note that the critical Mason number can be approximated when the ratio becomes unity, $U^{DIP}_{rms}/U^{Shear}_{rms} = 1$.}\label{fig:fig7}
\end{figure}

Here, we focus more on the stripped pattern observed in figure~\ref{fig:fig5}(\textit{b}) at $Mn = 1$ and figure~\ref{fig:fig5}(\textit{c}) at $Mn = 0.1$. It is known that the particles tend to pair up briefly due to DIP or ICEP along the field ($y$) direction and then immediately reorient toward the directions orthogonal to the field direction, after which they tend to repel from each other \citep{saintillan2008,park2011}. Under sufficiently strong particle loading, specifically beyond the semi-dilute regime ($\phi > 15\%$), we previously found that such pairing dynamics cause the particles to interact dominantly along the transverse directions ($x,z$), resulting in the maximum probability of the pair distribution function at the particle equators \citep{mirpark2019}, which is also seen in figure~\ref{fig:fig5}. Upon the application of shear flow at which the shear effect is comparable to or weaker than the DIP effect, the particles tend to be trapped in the flow-vorticity ($x$-$z$) plane and thus self-assemble into two-dimensional horizontal sheets while still coherently moving in the flow ($x$) direction. These two-dimensional sheet-like structures appear to experience no direct contact between them, making them move more easily (see accompanying supplementary movies for $Mn = 1$ at $\phi = 30\%$ and $Mn = 0.1$ at $\phi = 45\%$). Such a well-organized structural pattern under the flow is associated with less energy dissipation, resulting in a shear viscosity smaller than for the absence of the electric field, as seen in figure~\ref{fig:fig3}(\textit{b}). However, these sheet-like structures are gradually disrupted by further increasing the Mason number, forming hydroclusters and, in turn, leading to the viscosity increase.

\subsection{Suspension dynamics}
We now attempt to illuminate the transition observed in the suspension rheology and microstructure around $Mn = 1$ from a suspension dynamics point of view. A straightforward way to examine suspension dynamics is kinetics. We compute the contributions of DIP and shear to the root-mean-square (RMS) velocity of particles, which are denoted as $U^{DIP}_{rms}$ and $U^{Shear}_{rms}$, respectively. Figure~\ref{fig:fig7} presents the ratio of these two contributions, $U^{DIP}_{rms}/U^{Shear}_{rms}$, as a function of the Mason number for different volume fractions. The ratio seems to decrease almost linearly on the logarithmic scales with the Mason number. It passes the unity at $Mn = 0.7 - 1$, around which there would be a changeover in the dominant mechanism of particle kinetics from DIP to shear dominance. Such a changeover could confirm the critical Mason number ($Mn \approx 1$), which is identified by the relative shear viscosity in figure~\ref{fig:fig3}(\textit{b}), the normal stress differences in figure~\ref{fig:fig4}, and the microstructure in figure~\ref{fig:fig5}. The inset of figure~\ref{fig:fig7} shows the changeover Mason number $Mn_C$ as a function of volume fraction. There seems to be no general trend but in the range of $Mn_C = 0.7 - 1$. Similar to the normal stress differences, large errors are observed at higher volume fractions due to stronger fluctuations. It is worth noting again that even small errors in the computation of each contribution of DIP and shear to the RMS velocity can give rise to larger errors in the calculation of their ratio, which is also noted by \cite{sierou2002}.

\begin{figure}
  \centerline{
  \includegraphics[width=5.0 in]{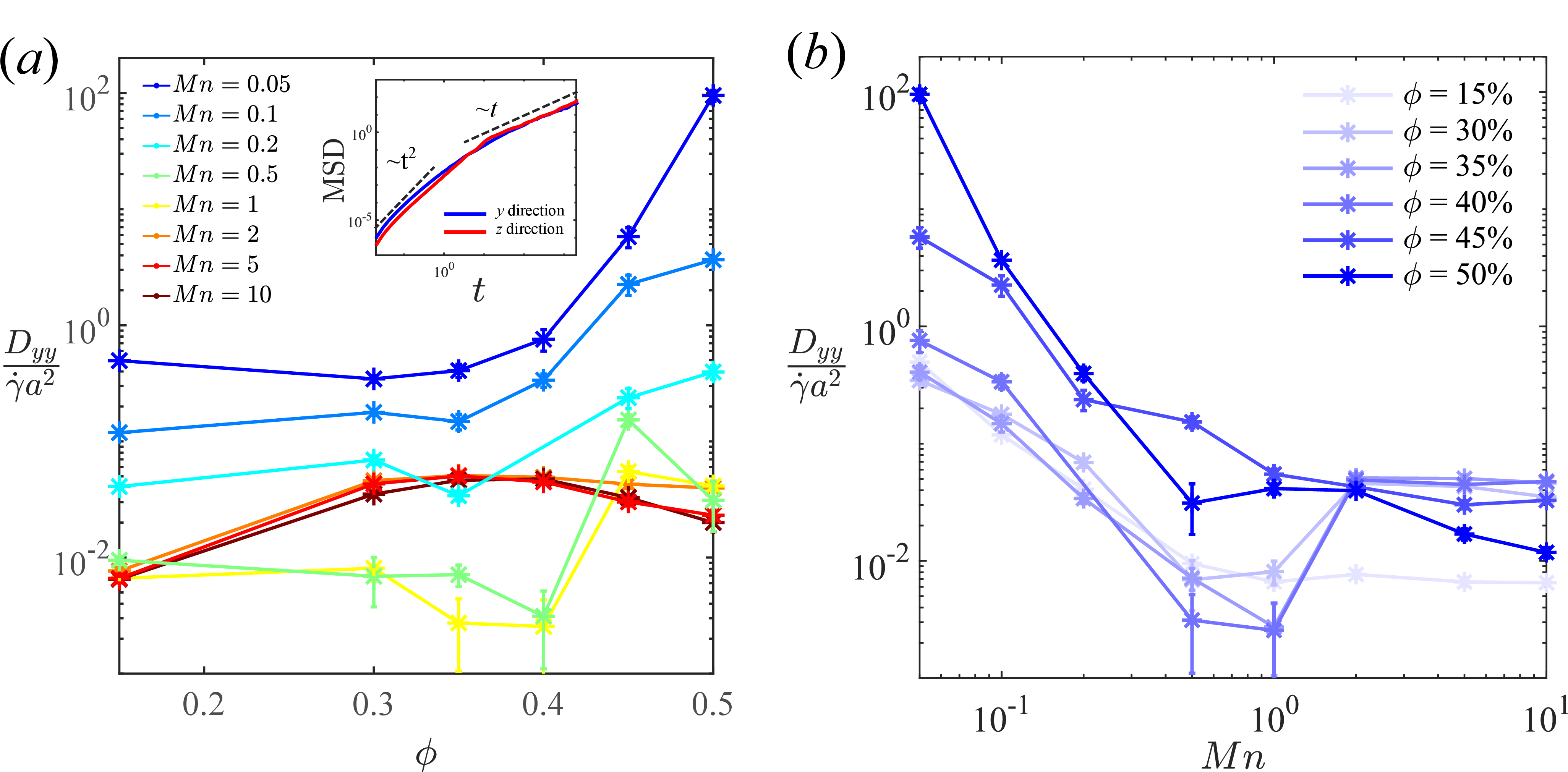}}
  \caption{(\textit{a}) Hydrodynamic diffusivity $D_{yy}$ in the electric-field ($y$) direction, which is the same as the velocity-gradient direction of the external flow, as a function of volume fraction for a range of Mason numbers. Inset: the mean-square displacement (MSD) curves in the velocity-gradient ($y$) and vorticity ($z$) directions at $\phi = 30\%$ and $Mn = 0.05$. (\textit{b}) The diffusivity is replotted as a function of Mason number for various volume fractions.}\label{fig:fig6}
\end{figure}

Finally, we investigate the hydrodynamic diffusion to make a further link between the suspension dynamics and the transition observed in the rheological properties and microstructure. To effectively quantify the hydrodynamic diffusion of a suspension, the mean-square displacements (MSD) over time are calculated, as seen in the inset of figure~\ref{fig:fig6}(\textit{a}) as an example in the log-log plot. The MSD curve exhibits an initial quadratic growth with time followed by the diffusive regime with linear growth due to particle-particle interactions. The average slopes of the MSD curves over the diffusive regime give the effective hydrodynamic diffusivity tensor ${\mathsfbi{D}}$ \citep{park2010,mirpark2019}. Figure~\ref{fig:fig6}(\textit{a}) shows the dependence of the diffusivity in the velocity-gradient direction $D_{yy}$ on the volume fraction for various Mason numbers. $D_{yy}$ also denotes the hydrodynamic diffusivity in the field direction. Note that the diffusivity $D_{yy}$ is normalized by $\dot{\gamma}a^2$. At the smallest Mason number of $Mn = 0.05$, the diffusivity appears to be the largest across all volume fractions considered due to the strongest DIP interactions in comparison to other higher Mason numbers. At this Mason number, the diffusivity remains almost constant up to $\phi \approx 35\%$, after which it starts to increase with volume fraction. As the Mason number is increased, this trend seems to maintain up to $Mn = 1$, but its magnitudes continue to decrease and reach the minimum at $Mn = 1$. 
Interestingly, as the Mason number is further increased, the diffusivity starts to increase, especially for $20\% < \phi < 45\%$. It then seems to collapse on a single curve for $Mn \geqslant 5$ as it approaches the pure-shear-flow diffusivity, which is comparable to one observed by \cite{sierou2004}. To more clearly demonstrate the transition around $Mn = 1$, the normalized diffusivity is replotted in figure~\ref{fig:fig6}(\textit{b}) as a function of the Mason number for various volume fractions. As the Mason number increases, the diffusivity continues to decrease until $Mn \approx 1$, after which it becomes almost constant, except for $\phi = 35\%$ and $40\%$, where there is a slight increase from $Mn = 1$ to $Mn = 2$. Nevertheless, a clear transition is observed around $Mn = 1$ in figure~\ref{fig:fig6}($b$). Therefore, the suspension dynamics, characterized by the kinetics and hydrodynamic diffusion in figures~\ref{fig:fig7} and \ref{fig:fig6}, strongly supports a changeover in the dominant mechanism around the critical Mason number of $Mn = 1$, which ties closely into the transition observed in rheological properties and microstructure around $Mn = 1$.

\section{Conclusion}\label{sec:conclude3}
We have investigated for the first time the effects of dipolophoresis (the combination of two nonlinear electrokinetic phenomena -- dielectrophoresis and induced-charge electrophoresis) on the shear rheology of suspensions of ideally conductive spheres using large-scale numerical simulations. For ideally conductive particles undergoing dipolophoresis (DIP) in an electric field, the induced-charge electrophoresis (ICEP) is known to be predominant over the dielectrophoresis (DEP) -- thus DEP effect is mostly neglected. To simulate the particle motions driven by both dipolophoresis and shear flow, we developed a simulation model that incorporates our previous model for dipolophoresis using the mobility-based Stokesian dynamics approach \citep{mirpark2019} combined with the classical resistance-based Stokesian dynamics approach \citep{brady1988} for shear flow in a periodic domain of the Lees-Edwards kind. The suspension is characterized by Mason number ($Mn$), which is the ratio of viscous shear stress to electrokinetics-driven stress, $Mn=\eta \dot{\gamma}/ \varepsilon E_0^2$. The suspension rheology was examined in a range of $5 \times 10^{-2} \le Mn \le 10^2$ at volume fractions up to $\phi = 50\%$.

For shear viscosity, it was found that at high shear rates corresponding to high Mason numbers ($Mn > 5$), the $\phi$-dependent shear viscosities collapse almost onto a single curve of a purely sheared suspension, demonstrating almost negligible effects of dipolophoresis. As the Mason number is decreased below $Mn = 5$, the $\phi$-dependent shear viscosity starts to decrease due to the increasing DIP effect, where a significant reduction of the viscosity is observed from $Mn = 2$ to $Mn = 1$. As the Mason number is further decreased below $Mn = 1$, the viscosity slightly decreases as entering the low-shear-rate regime, where the $\phi$-dependent viscosity eventually collapses again onto another single curve for $Mn < 0.5$. For shear-dependent behaviours, it was found that the shear viscosity at a volume fraction as small as $\phi = 15\%$ is insensitive to a shear rate, showing Newtonian-like behaviour. For the semi-dilute regime at $\phi \approx 30\%$, the shear thickening starts to occur in which viscosity starts to sharply increase around $Mn = 1$. In the concentrated regime of $\phi = 45\%-50\%$, an almost five-fold increase in viscosity occurs during the shear thickening process. The microstructure was also investigated, where for low shear rates or DIP-dominant regimes $Mn \le 1$ and beyond semi-dilute regimes $\phi \ge 30\%$, the two-dimensional sheet-like structures are formed in the plane orthogonal to the velocity-gradient direction or electric-field direction. This structure helps a suspension flow easily, which is responsible for low viscosity. As a shear rate or the Mason number is increased above $Mn \approx 1$, such an organized structure is gradually disrupted, leading to the formation of the so-called hydroclusters, which eventually results in increasing viscosity and thus promoting the shear thickening.
  
For the normal stress differences, it was found that at low Mason numbers below $Mn \approx 1$, the first normal stress difference $N_1$ is positive and the second normal stress difference $N_2$ is negative at a low volume fraction of $\phi  = 15\%$, implying that the effects of DIP on the normal stress differences are reminiscent of the Brownian contribution \citep{foss2000}. As a volume fraction is further increased for $\phi \geqslant 30\%$, $N_1$ and $N_2$ show the opposite trend that $N_1$ and $N_2$ become negative and positive, respectively. This can be explained by the microstructure that exhibits the high probability of the pair distribution function in the flow and vorticity directions due to the effect of DIP, causing the particles to get trapped in two-dimensional sheets orthogonal to the velocity-gradient direction. For higher Mason numbers of $Mn > 1$, both $N_1$ and $N_2$ become negative across all volume fractions considered, and their magnitude increase with volume fraction.
 
To further illustrate the transition observed in rheological properties around $Mn = 1$ for which we call the critical Mason number, suspension dynamics were investigated. The particle kinetics of each DIP and shear contribution is compared, showing that a changeover in the dominant kinetics is observed around $Mn = 0.7 - 1$ from DIP to shear dominance. Furthermore, the hydrodynamic diffusivity in the velocity-gradient direction clearly exhibits the transition around $Mn = 1$. Its magnitude is the largest at the lowest Mason numbers $Mn = 0.05$ due to the strong DIP effect and starts to decrease until reaching a minimum at $Mn \approx 1$, beyond which it becomes almost constant.

Lastly, the present study suggests the potential use of DIP or ICEP as a means to control the suspension rheology of conductive particles in an electric field. For instance, at high volume fractions (i.e., $\phi \ge 30\%$) and high shear rates (i.e., $Mn > 1$), having a stronger DIP effect via increasing the electric-field strength could reduce the viscosity from shear-thickened states. \SM{In addition, it also suggests that the direction or frequency of an electric field could play a promising role as a control parameter in controlling the rheology of such suspensions, allowing to tune shear-thinning or shear-thickening behaviours. A detailed investigation of the active rheology control via an electric field for suspensions of conductive particles will be a subject of interesting future work.} 

\section*{Acknowledgements}
The authors gratefully acknowledge the financial support from the National Science Foundation through grants CBET-1936065 and CBET-2154788 and the Collaboration Initiative at the University of Nebraska. This work was completed utilizing the Holland Computing Center of the University of Nebraska, which receives support from the Nebraska Research Initiative.

\section*{Declaration of Interests}
The authors report no conflict of interest.
 
\bibliographystyle{jfm}
\bibliography{JFM_shear_test}

\end{document}